\documentclass[11pt]{article}
\usepackage{amscd,amssymb,amsmath,latexsym,enumerate,bm}

\newtheorem{theorem}{Theorem}[section]
\newtheorem{lemma}[theorem]{Lemma}

\begin{document}

\title{On the Generalized Wannier Functions}
\author{Emil Prodan \\
Department of Physics, Yeshiva University, New York, NY 10016} \maketitle
\begin{abstract}
We consider single particle Schrodinger operators with a gap in the
energy spectrum. We construct a complete, orthonormal basis function set for
the invariant space corresponding to the spectrum below the spectral gap, which are exponentially localized around a set of closed surfaces of monotonically increasing sizes. Estimates on the exponential decay rate and a
discussion of the geometry of these surfaces is included.
\end{abstract}


\section{Introduction}

The fast developing field of nano-science and the need for
microscopic understanding of the biological processes are just two
of the driving forces for the search of computationally efficient
electronic structure algorithms. One important goal is to develop
electronic structure algorithms that scale linearly with the number
of particles. It seems that such algorithms will involve a
localized, real-space description of the electronic structure
\cite{Goedecker} and, of course, Wannier functions will play a major
role \cite{Kohn73,Kohn74}.

The Wannier functions were originally defined for perfectly periodic
insulating crystals \cite{Wannier}, {\emph i.e.} those with a gap in the energy spectrum and with Fermi energy pinned in the middle of this gap. The key property of the Wannier functions is  that they form a complete basis of uniformly exponentially localized functions (relative to the Wannier centers) for the invariant subspace corresponding to the energy spectrum below the gap. The latter will be referred to as the space of occupied electron states. This property has been proven for 1D periodic systems \cite{Kohn59}, molecular chains with inversion symmetry \cite{Prodan} and for simple bands in more than 1 dimension \cite{Cloizeaux1,Cloizeaux2,Nenciu83}. A natural and important question is if one can define the equivalent of the Wannier functions for systems that are not periodic. Examples of such generalized Wannier functions were given by Kohn and Onffroy \cite{KohnOnffroy} for crystals with one impurity and Rehr and Kohn \cite{KohnRehr} for crystals with surfaces. In a relatively recent
paper \cite{Nenciu98}, Nenciu conjectured that any system with a gap
in the energy spectrum should posses exponentially localized Wannier
functions and he proved the conjecture in one dimension. More precisely, according to this conjecture, it is expected that there exists a discrete set $\Gamma$ of $\mathbb R^d$ and a set of generalized Wannier functions, $\big \{\psi_{\vec g,j}(\vec r)\big \}^{\vec g \in \Gamma}_{1\leq j \leq m(\vec g) < \infty}$, forming a complete basis for the space of occupied electron states and, moreover,
\begin{equation}\label{Exp1}
\int_{\mathbb R^d} d\vec r \ e^{2\alpha|\vec r - \vec g|} |\psi_{\vec g,j}(\vec r)|^2 \leq M < \infty,
\end{equation}
where $\alpha$ and $M$ can be chosen independent of the indices, hence providing a uniform bound. The proof by Nenciu in dimension 1 \cite{Nenciu98}
is based on an idea introduced by Kivelson \cite{Kivelson}, who
showed that, for 1D perfectly periodic systems, the eigenvectors of
$P_0xP_0$ provide a set of Wannier functions ($P_0=$ the band
projection operator). Moreover, it was later shown by Marzari and
Vanderbilt \cite{Vanderbilt} that these Wannier functions are
maximally localized. Further important results were established in \cite{Cornean} for quasi 1-dimensional systems. Now, if one tries to extend Kivelson's
construction to more than one dimension, the problem is that, even
for perfectly periodic systems, $P_0xP_0$, $P_0yP_0$ and $P_0zP_0$
do not commute, and any of these three operators have, in general,
continuum spectrum and, consequently, no localized eigenvectors.

In this paper, we propose an alternative way of looking into the
problem by replacing the discrete set $\Gamma$ with a collection of spheres or of other closed surfaces. More precisely, let us consider a
$\vec{\nabla}^2$-bounded potential $v$ over ${\mathbb R}^d$
($d=1,2,\ldots$), with relative bound less than one. We consider
only potentials that are bounded from below. Measuring the energy
from the bottom of the potential, it is equivalent to say that the
potentials are positive. Other than this, the only assumption on $v$
is that the self-adjoint Hamiltonian,
\begin{equation}
    H:{\mathcal D}\big (-\vec{\nabla}^2 \big )\rightarrow L^2 \big ({\mathbb R}^d \big ), \
    \ H=-\vec{\nabla}^2+v,
\end{equation}
has a gap in the energy spectrum. Let ${\mathcal K = P_0 L^2 \big ({\mathbb R}^d \big ) }$ be the invariant space corresponding to the spectrum below the gap. In these conditions, the following statements hold.\bigskip

\begin{theorem}[Main Result]\label{MainTh} For $q>0$, let $W_{q}$ be the
following bounded, self-adjoint operator
\begin{equation}
 W_q:{\mathcal K}\rightarrow {\mathcal K}, \ \ W_q = P_0e^{-q\rho}P_0,
\end{equation}
where $\rho =\sqrt{1+r^2}$. Then:

\begin{enumerate}[\rm (i)]

\item The spectrum of $W_q$, $\sigma (W_q)$, is discrete, of finite degeneracy, and has one and only one accumulation at zero.

\item All eigenvectors of $W_q$ decay exponentially as $|\vec{r}|\rightarrow \infty$, with a rate $\alpha$ which larger than or equal to $q$. Furthermore, the optimal exponential decay rate method which can be produced by the present accepts a lower bound which can be computed entirely from the spectrum.

\item Let $\Big \{ \lambda_i,\big \{ \psi_{i,j}(\vec{r}) \big \}_{1\leq j \leq m_i < \infty}\Big \}_{i \in \mathbb N}$ be the set of decreasingly ordered eigenvalues (of degeneracy $m_i$) and their corresponding system of eigenvectors. Define
\begin{equation}\label{radius}
    \rho_i=\sqrt{\left( \frac{\ln \lambda _{i}}{q}\right)^2-1}.
\end{equation}
Then
\begin{equation}\label{Exp2}
    \int_{\mathbb R^d} e^{2\alpha |\rho-\rho_i|'}|\psi_{i,j}(\vec{r})|^2 d\vec{r}\leqslant
    M < \infty,
\end{equation}
with $M$ independent of the indices and $|x|'=x$ if $x\geq0$ and $|x|'=-\frac{1}{2}x$ if $x < 0$.

 \end{enumerate}

\end{theorem}

Several remarks are in order: 

\begin{enumerate}[\rm R1)]

\item The reader should notice the similarity between Eqs.~\ref{Exp1} and \ref{Exp2} and of the conditions in which the two statements apply. On these grounds, the set of functions defined above can be rightfully regarded as generalized Wannier functions.

\item Apart from the fact that the radii $\rho_i$ are monotonically increasing to $\infty$, we don't have much of a characterization of the set of spheres. For example, we cannot rule out the scenario in which the spacing $\rho_{i+1}-\rho_i$ between the spheres decreases as $i^{-1}$ (and $\rho_i$ still increase monotonically to $\infty$). Note, however, that the characterization of the set $\Gamma$ is similarly coarse in \cite{Nenciu98}, and similar scenarios could apply to it.  

\item Since $W_q$ is self-adjoint and $\sigma(W_q)$ is discrete and of finite degeneracy, its eigenvectors form a complete and orthonormal basis for ${\mathcal K}$. Thus, simply by solving for the eigenvalues of $W_q$, one can
construct a complete, orthonormal and uniformly exponentially decaying set of basis functions for the subspace ${\mathcal K}$. 

\item The matrix elements of the Hamiltonian in the generalized  Wannier basis decay at least as
\begin{equation}\label{overlap}
\big \langle \psi_{i,j},H \psi_{i',j'} \big \rangle \leq  {\rm ct.} \, e^{-\alpha|\rho_i - \rho_{i'}|},
\end{equation}
where the constant is uniform. Then, provided the radii $\rho_i$ increase sufficiently fast, the generalized Wannier basis can be used to generate accurate short range tight-binding Hamiltonians.

 \item By simply modifying the function form of $\rho$, we can obtain
generalized Wannier functions that are localized around more general
closed surfaces like ellipsoids, etc., which may come in handy when defects of crystals are investigated.

\end{enumerate}

\section{Proof of the main statement}

The proof of our main statement involves the analytic continuation
technique, first used by des Cloizeaux \cite{Cloizeaux1,Cloizeaux2}
in the context of band calculations. Let $P_0$ denote the spectral projection onto the states below the gap. Let $E_{\pm }$ denote the
upper respectively the lower edge of the gap $\Delta $ and consider
the unitary transformation $e^{-iq\rho }$, with $q$ real. We define
the following unitarily equivalent representations of the spectral 
projection,
\begin{equation}
    P_{0}(q) =e^{-iq\rho }P_{0}e^{iq\rho }.
\end{equation}
Throughout, $\| \cdot \|$ will denote the operator norm $\|A\|=\sup_{\|\psi\|=1}\sqrt{\langle A\psi, A\psi
\rangle}$. We will show later that we can analytically extend the family $P_0(q)$ to complex $q$. More precisely:

\begin{lemma}\label{Lemma1} $P_{0}\left( q\right) $ extends to an
analytic family of bounded operators,
\begin{equation}
    \|P_0(q)\| < \infty,
\end{equation}
for any $q$ in a complex strip $\left| {\rm Im} \, q\right| <q_{M}$. Moreover, $q_{M}$ is always larger than or equal to
\begin{equation}\label{qcritic}
    q_c=q_0\left( \sqrt{1+\Delta /q_{0}^{2}}-1\right),
\end{equation}
where $q_0=\sqrt{E_{-}}+\sqrt{E_{+}}$.
\end{lemma}

We start now the proof of Theorem~\ref{MainTh}. (i) We write $W_q$ as
\begin{equation}
    W_q=P_0\int_{\gamma }e^{-q\rho }(z-H)^{-1}\frac{dz}{2\pi i},
\end{equation}
with $\gamma$ a contour surrounding the spectrum below the gap. Let $a$ be a positive constant. Since the functions $e^{-q\sqrt{1+r^2}}$ and $(r^2+a)^{-1}$ are in $L^s(\mathbb R^d)$ if $s$ is taken sufficiently large, then $e^{-q\rho}(-\vec \nabla^2+a)^{-1}$ belongs to the $s$-Schatten class \cite{Simon1}. Furthermore, since
$(-\vec \nabla^2+a)(z-H)^{-1}$ is bounded, it follows that $W_{q}$ is
in the $s$-Schatten class and, as a consequence, compact. The statement (i) then follows.

(ii) Let $\psi $ be a normalized eigenvector, $W_{q}\psi
=\lambda \psi $. Note that $P_0\psi=\psi$. Since
\begin{equation}\label{eigenvalue}
    \lambda =\langle \psi, W_q \psi \rangle = \int_{\mathbb R^d} e^{-q\rho }| \psi (\vec{r}) |^2d\vec{r}
\end{equation}
and $\rho\geq 1$, all eigenvalues are strictly positive and smaller
than $e^{-q}$. Let us now consider  a $q$ smaller than the $q_M$ of
Lemma~\ref{Lemma1}. We will prove
\begin{equation}
    \int e^{2q\rho}|\psi(\vec{r})|^2d\vec{r}<\infty,
\end{equation}
which clearly shows that $|\psi(\vec{r})|$ decays faster than
$e^{-q\rho}$ as $|\vec{r}|\rightarrow \infty$. Indeed, observing
that $e^{q\rho }W_q=P_0(iq)P_0$, we find
\begin{eqnarray}\label{qless}
    \left[\int e^{2q\rho}|\psi(\vec{r}|^2d\vec{r}\right]^{1/2}&=&
    \|e^{q\rho}\psi\|=\lambda^{-1}\|e^{q\rho}W_q\psi\| \\
    &=&\lambda^{-1}\|P_0(iq)\psi\| \leqslant
    \lambda^{-1}\|P_0(iq)\|,\nonumber
\end{eqnarray}
and, by definition, $q_{M}$ is the maximum value of $q$ for which
$\| P_0(iq)\|< \infty$. The conclusion is that all eigenfunctions of $W_q$ decay exponentially as $r \rightarrow \infty$, with a rate $\alpha = q$. By optimization, this rate can be made larger than or equal to the value given in Eq.~\ref{qcritic}. Note, however, that the constant in front of the exponentially decay estimate is $\lambda^{-1}$, which is not uniform because $\lambda$'s accumulate at zero. As such, the next point of the Theorem is crucial.

(iii) Let us write the eigenvalues of $W_q$ as
$\lambda_i=e^{-q\rho_i}$. Since $\lambda_i$ are positive and
accumulate at zero, $\left\{ \rho_i\right\}_i$ is a sequence of
positive numbers which increase monotonically to infinity. If $\psi_i$ is an
eigenvector corresponding to $\lambda_i$, then using Eq.~(\ref{qless})
\begin{equation}
 \int e^{2q(\rho-\rho_i)}|\psi_i(\vec{r})|^2 d\vec{r}\leqslant
    \|P_0(iq)\|^2 .
\end{equation}
On the other hand
\begin{equation}
1=\lambda_i^{-1} \langle \psi_i, W_q \psi_i \rangle = \int e^{ q(\rho_i-\rho)}|\psi_i(\vec{r})|^2 d\vec{r}.
\end{equation}
By taking $M = \max\{1,\|P_0(iq)\|^2\}$, the estimate in Eq.~\ref{Exp2} follows. This concludes the proof of Theorem~\ref{MainTh}.

\vspace{0.2cm}

The rest of the paper contains the proof of Lemma~\ref{Lemma1}, which provides a lower bound for $q_M$. An important outcome is that this lower bound can be easily estimated from the band structure.

\section{Proof of Lemma~\ref{Lemma1}}

The proof follows Barbaroux et al \cite{Hislop}. We will try to
eliminate the unnecessary constants and optimize the
technique.\bigskip

\noindent \textbf{Proposition}. Consider $A$ and $B$, two bounded,
self-adjoint operators. Assume that $A$ has a spectral gap located
at $0$,
\begin{equation}
    d_\pm \equiv dist(\sigma_\pm(A) ,0) >0,
\end{equation}
where $\sigma_\pm$ denote the parts of the spectrum located above
and below zero. Let $P_\pm$ denote the spectral projector onto the
states corresponding to $\sigma_\pm(A)$ and $C>0$ be a bounded
self-adjoint operator such that $s_{+}\equiv d_+-\|P_+CP_+\|>0$.
Then, for $q$ real,
\begin{equation}\label{principal}
    \left\|(A-C+iqB)^{-1}\right\| \leqslant \left( 1-\frac{\left|
    q\right| }{\tilde{q}}\right)^{-1}\max \left\{\frac{1}{s_+},\frac{1}{d_-}\right\},
\end{equation}
where
\begin{equation}
    \tilde{q}=\frac{\sqrt{s_+d_-}}{\left\| P_{+}BP_{-}\right\|}.
\end{equation}
\bigskip

\noindent \textit{Proof.} Let $\varphi$ be a norm one but otherwise
arbitrary vector, $\varphi_\pm=P_\pm\varphi$ and $B_{+-}=P_+BP_-$.
We start from the observation
\begin{equation}\label{real}
    \text{Re}\left\langle \varphi_+-\varphi_-,(A-C+iqB)
    \varphi \right\rangle \leqslant \left\| (A-C+iqB) \varphi
    \right\|.
\end{equation}
The left hand side is equal to,
\begin{equation}
    \left\langle \varphi_+,(A-C)\varphi_+\right\rangle
    -\left\langle \varphi_-,(A-C) \varphi_-\right\rangle -2q
    \text{Im}\left\langle \varphi_+,B\varphi_-\right\rangle,
\end{equation}
and we have successively:
\begin{eqnarray}
    &&\|(A-C+iqB) \varphi \| \\
    &\geq& s_+ \left\| \varphi_+\right\|^2+d_-\left\| \varphi_-\right\|^2
    -2\left| q\right| \left\| B_{+-}\right\| \left\| \varphi
    _{+}\right\| \left\| \varphi _{-}\right\|  \notag \\
    &=&\left( 1-\left| q\right| /\tilde{q}\right) \left( s_{+}\left\| \varphi
    _{+}\right\| ^{2}+d_{-}\left\| \varphi _{-}\right\| ^{2}\right)  \notag \\
    &&+\left| q\right| \left\| B_{+-}\right\| \left( \sqrt[4]{\tfrac{s_{+}}{d_{-}%
    }}\left\| \varphi _{+}\right\| -\sqrt[4]{\tfrac{d_{-}}{s_{+}}}\left\|
    \varphi _{-}\right\| \right) ^{2},  \notag \\
    &\geq& \left( 1-\left| q\right| /\tilde{q}\right) \min\{s_+,d_-\}
\end{eqnarray}
and the affirmation follows.$\blacksquare \medskip $

We define the following analytic family,
\begin{equation}
    H_q=e^{-iq\rho }He^{iq\rho }=H+q^{2}|\vec{\nabla}\rho |^{2}+
    q\left( \vec{\nabla}\rho \cdot \vec{p}+\vec{p}\cdot \vec{\nabla}\rho
    \right),
\end{equation}
where we notice that $|\vec{\nabla}\rho |\leqslant 1$. Since
$\vec{\nabla}\rho \cdot \vec{p}+\vec{p}\cdot \vec{\nabla}\rho $ is
relatively $H$-bounded, $H_q$ extends to an analytic family of type
A over the entire complex plane \cite{Simon}. Then, according to
Combes and Thomas \cite{Combes},
\begin{equation}
    P_0(q)=\int_{\Gamma }(z-H_q)^{-1}\frac{dz}{2\pi i},
\end{equation}
where $\Gamma $ surrounds the lower band. One can see that $P_0(q)$
belongs to an analytic family of bounded operators as long as
$\Gamma$ belongs to the resolvent set of $H_q$. As we will argue
later, we have to worry only about the point where $\Gamma $ crosses
the real axis. We consider then an energy $E$ in the spectral gap
and calculate for which values of $q$ the resolvent $(H_q-E)^{-1}$
remains a bounded operator. It is enough to consider only purely
imaginary $q$. In the above Proposition, we take
\begin{equation}
    A=|H-E|^{-1/2}(H-E)|H-E|^{-1/2},
\end{equation}
\begin{equation}
    B=|H-E|^{-1/2}(\vec{\nabla}\rho \cdot \vec{p}+\vec{p} \cdot
    \vec{\nabla}\rho)| H-E|^{-1/2}
\end{equation}
and
\begin{equation}
    C=q^2|H-E|^{-1/2}|\vec{\nabla}\rho|^2|H-E|^{-1/2}.
\end{equation}
Then we have $d_-=1$ and $s_+=1-q^2/(E_+-E)$, and we evaluate
$\|B_{+-}\|$ from,
\begin{eqnarray}\label{Bpm}
    \|B_{+-}\| &=& \sup_{\varphi_+,\varphi_-}\left|
    \left\langle |H-E|^{-1/2}\varphi_+,\vec{\nabla}\rho \cdot
    \vec{p}|H-E|^{-1/2}\varphi_-\right\rangle \right. \\
    &&\left. +\left\langle \vec{\nabla}\rho \cdot \vec{p}|H-E|^{-1/2}
    \varphi_+,|H-E|^{-1/2}\varphi_-\right\rangle\right|,  \notag
\end{eqnarray}
where the supremum goes over all norm one $\varphi_+$ and
$\varphi_-$ in the upper respectively lower band. If
\begin{equation}
    \psi_{\pm }=(H+a)^{1/2}|E-H|^{-1/2}\varphi_{\pm},
\end{equation}
for some positive $a$, then we find
\begin{equation}
    \left\|B_{+-}\right\| \leq \left\| p(H+a)^{-1/2}\right\|
    \left(\frac{\left\| \psi_-\right\| }{\sqrt{E_{+}-E}}+\frac{\left\| \psi
    _+\right\| }{\sqrt{E-E_{-}}}\right).
\end{equation}
Since the potential is positive, $\left\| p(H+a)^{-1/2}\right\| \leq
1$ for all $a>0$. We take the limit $a=0$ and use the spectral
theorem to evaluate $\left\| \psi_{\pm }\right\| $ and finally find
\begin{equation}
    \left\| B_{+-}\right\| \leq \frac{\sqrt{E_-}+\sqrt{E_+}}
    {\sqrt{\left( E_+-E\right) \left(E-E_-\right) }}.
\end{equation}
From Eq.~(\ref{principal}), we can conclude
\begin{equation}
\left\| \left| H-E\right| ^{1/2}\left( H_{iq}-E\right) ^{-1}\left|
H-E\right| ^{1/2}\right\| \leqslant \frac{1}{s_{+}\left( 1-\left| q\right| /%
\tilde{q}\right) }\text{,}  \label{preliminary}
\end{equation}
where
\begin{equation}
\tilde{q}=\frac{\sqrt{\left( E-E_{-}\right) \left( E_{+}-E-q^{2}\right) }}{%
\sqrt{E_{-}}+\sqrt{E_{+}}}\text{.}
\end{equation}
The optimal position of the energy $E$ in the spectral gap for $\tilde{q}$
to become maximum is $E_{c}=(E_{-}+E_{+}-q^{2})/2$ which leads to $\tilde{q}%
=(\Delta -q^{2})/2q_{0}$. Then from Eq.~(\ref{preliminary}),
\begin{equation}
\left\| \left( H_{iq}-E_{c}\right) ^{-1}\right\| \leqslant \frac{\Delta
+q^{2}}{\Delta -q^{2}}\frac{2}{\Delta -q^{2}-2q_{0}\left| q\right| }\text{,}
\end{equation}
and the right side is finite as long as
\begin{equation}
q<q_{0}\left( \sqrt{1+\Delta /q_{0}^{2}}-1\right) \text{.}
\end{equation}
As one can easily see, adding an imaginary part to $E_{c}$ will not
affect the real part in Eq.~(\ref{real}) and the immediate
consequence of this is that $\left( H_{iq}-E_{c}+i\varepsilon
\right) ^{-1}$ is bounded as long as $\left( H_{iq}-E_{c}\right)
^{-1}$ is bounded. Consequently, for $\left| \text{Im}q\right|
<q_{c}$, we can always find a curve $\Gamma $ surrounding the lower
band and lying in the resolvent set of $H_{iq}$.$\blacksquare $

\section*{ACKNOWLEDGMENTS}
This work was supported by the U.S. NSF grant DMR-1056168.

\end{document}